\documentstyle[preprint,aps]{revtex}
\begin{document}
\tightenlines
\preprint{Preprint CBPF-NF-012/02, to appear in 
Braz. J. Phys. {\bf 33}, 1 (2003).}
\title{A REPRESENTATION OF THE VIRASORO ALGEBRA
VIA WIGNER-HEISENBERG ALGEBRAIC TECHNIQUE TO BOSONIC
SYSTEMS}
\author {E. L. da Gra\c{c}a$^{\dagger}$, H. L. Carrion$^{\ddagger}$ and
R. de Lima Rodrigues$^{\alpha}$\thanks{Permanent address: Departamento 
de Ci\^encias Exatas e da Natureza,
Centro de Forma\c{c}\~ao de Professores, Universidade Federal de Campina Grande,
Cajazeiras -- PB, 58.900-000, Brazil. E-mail rafaelr@cbpf.br or 
rafael@fisica.ufpb.br. E-mail for H. L. Carrion is  hleny@cbpf.br.} \\
${}^{\ddagger,\dagger,\alpha}$ Centro Brasileiro de Pesquisas F\'\i sicas\\
Rua Dr. Xavier Sigaud, 150\\
CEP 22290-180, Rio de Janeiro-RJ, Brazil\\
${}^{\dagger}$ Departamento de F\'\i sica,
Universidade Federal Rural do Rio de Janeiro\\
Antiga Rodovia Rio-S\~ao Paulo Km 47, BR 465,
CEP 23.890-000, Serop\'edica-RJ.
}

\maketitle

\begin{abstract}
  Using the Wigner-Heisenberg algebra for bosonic systems in connection with
oscillators we find a new representation for the Virasoro algebra.
\end{abstract}

\vspace{0.5cm}


\pacs
\newpage

\section{Introduction}

In 1950, Wigner\cite{Wigner50}, proposed the interesting question, 
"Do the equations of motion determine the quantum-mechanical commutation 
relations?" and found as answer a generalized quantum rule for the 
one-dimensional harmonic oscillator. 
In the next year, Yang \cite{Yang} found the coordinate representation 
for the linear momentum operator.
Yang's wave mechanical description was further studied by Ohnuki et al. 
\cite{Ohnuki} and Mukunda et al. \cite{Sharma}.
Recently, the general Wigner-Heisenberg (WH) oscillator algebra 
\cite{Green,Ryan,Deser,JR90} has been investigated in the context of the 
deformed algebra \cite{Mi97}. There, the author
shows that finite-dimensional representations of the deformed 
parafermionic algebra with internal $Z_{2}$-grading structure.
The superealization of the  WH algebra has been independently considered in two
works \cite{JR90,Mi00}.

 The Virasoro algebra has been several applications in literature, 
let us point, for stance, the connections with the
conformal group 
\cite{Top}, construction as
$su(1,1)$ extension   \cite{zac87}, 
supervirasoro \cite{zac88} and
quantum algebras
\cite{zac90}.

In this work, starting from the  Wigner-Heisenberg algebraic
technique for the  bosonic systems in connection with general
oscillator, we find a new representation for the Virasoro algebra.

This work is arranged in the following way. In Section II, we present the WH
algebra. In Section III, a representation of the modified Virasoro 
algebra is found. The conclusions are drawn in the Section IV.

\section{THE WH ALGEBRA}

The Wigner Hamiltonian expressed in the symmetrized bilinear form 
in terms of the mutually adjoint abstract operators $\hat a{^\pm},$ defined by

\begin{equation}
\label{w1} 
\hat H{_W} = \frac{1}{2}( \hat {p}^{2}_x+ \hat x^{2} ) =
\frac{1}{2}[\hat a^{-}, \hat a^{+}]_+ = \frac{1}{2}(\hat a^{-} \hat a^{+} +
\hat a^{+} \hat a^{-}),
\end{equation}
where

\begin{equation}
\label{w2}
\hat {a}^{\pm} = \frac{1}{\sqrt {2}} (\pm i\hat {p}_x - \hat x).
\end{equation}
Wigner showed that Heisenberg's equations of motion

\begin{equation}
\label{w3}
[\hat {H}_{W}, \hat {a}^{\pm}]_- = \pm \hat {a}^{\pm},
\end{equation}
do not necessarily entail in the usual quantum rule 

\begin{equation}
\label{RCC} [a^{-}, a^{+}]_{-} = 1 \Rightarrow [\hat x,\hat p_{x}]_{-} =i, 
\quad \hbar=1,
\end{equation}
but a more general quantum rule \cite{Yang,Ohnuki,Sharma} given by

\begin{equation}
\label{CW}
[\hat a{^-}, \hat a{^+}]_- = 1 + c \hat R \Longrightarrow
[\hat {x}, \hat {p}_x]_- = i (1 + c \hat {R}),
\end{equation}
where $c$ is a real constant, related to the ground state energy 
$E^{(0)}_W \geq 0$ by virtue of the positive semi-definite form 
of $\hat H_W$\footnote{Note that the case $c=0$ corresponds 
to the usual oscillator with $E^{(0)}=\frac{1}{2}, \quad \hbar=\omega=1.$}

\begin{equation}
\label{w5}
\vert c \vert = 2E^{(0)} - 1,
\end{equation}
which is called Wigner parameter.

The basic (anti-)commutation relations (\ref{w1}) and
(\ref{w3}), together with the derived relation (\ref{CW}), 
are referred to as constituting the WH algebra. It is a parabose algebra 
\cite{Green} for the degree of freedom.
The WH algebra is obtained by combining also the requeriment that 
$\hat x$ satisfies the classical equation of motion 
($\ddot{\hat x} + \hat x = 0$).

Note that $\hat R$ is an abstract operator satisfying the properties
 
\begin{equation}
\label{w6}
[\hat R, \hat {a}^{\pm}]_+ = 0 \Rightarrow [\hat {R}, \hat{H}_{W}]_- = 0;
\quad \hat R^{\dagger} = \hat {R}^{-1} = \hat R, \quad \hat {R}^2 = 1,
\end{equation}
where one has used the following notation for the 
(anti-)commutation relations: $[A,B]_+\equiv
AB+BA$ and $[A,B]_-\equiv AB-BA.$ Besides, we have

\begin{eqnarray}
\label{w7}
H_W &=& \hat{a}^{+}\hat{a}^{-} + \frac{1}{2}(1 + c \hat{R})\nonumber\\
&=& \hat{a}^{-}\hat{a}^{+} - \frac{1}{2}(1 + c \hat{R}).
\end{eqnarray}

 Abstractly (O'Raifeartaigh and Ryan \cite{Ryan}, Boulware and  Deser
\cite{Deser}) $\hat R$
is the Klein operator, $\pm \hbox{exp}\{i\pi(\hat H_{W} - E^{(0)}_W)\}.$

In the mechanical representation first investigated by
Yang\cite{Yang}, $\hat R$ is realized by the Parity operator $P:$

\begin{equation}
\label{w8}
P|x> = |-x> \Rightarrow P x P^{-1} = -x, \quad P p_x P^{-1} = -p_x, \quad
P^{-1} = P, \quad P^2=1.
\end{equation}
Indeed, Yang \cite{Yang} found the coordinate representation 
for the momentum operator $p_x$ as given by

\begin{equation}
\label{w9}
\hat p{_x} \longrightarrow p{_x} = -i \frac{d}{dx} + i\frac{c}{2x}P, \quad
\hat x \longrightarrow x,
\end{equation}

\begin{equation}
\hat a^{\pm} \longrightarrow 
a^{\pm}_{\frac c2}=\frac{1}{\sqrt {2}}
\left(\pm \frac{d}{dx} \mp \frac{c}{2x}P - x \right).
\end{equation}
Yang's wave mechanical description was further investigated 
in \cite{Ohnuki,Sharma}.

\section{ Modified Virasoro Algebra}

The elements of the Virasoro algebra \cite{Top,zac88}
will be represented in terms of ladder operators $a^{\pm}$
of quantum oscillators with or without deformation.

\subsection{Canonical representation of the Virasoro algebra}

The conformal group, $G,$ in 2 dimensions consists of all general
transformations:

\begin{equation}
\Gamma : z \rightarrow \xi(z), \quad \bar\Gamma: \bar{z} \rightarrow
\bar{\xi}(\bar{z}),
\end{equation}
where $\Gamma$ is a group of the more general transformation with
one coordinate, or equivalently is a group of the 
{\it diffeomorphism transformations} in one dimension. The same
situation for $\bar{\Gamma}.$ Then, $G$ is given by the direct
product, viz.,

\begin{equation}
G = \Gamma  \otimes  \bar{\Gamma}.
\end{equation}

In the literature, $\Gamma$ and $\bar{\Gamma}$ are
usually referred to as  chiral and anti-chiral components of the 
conformal algebra in 2-dimensions. The  algebra associated to the $\Gamma $ is
called {\it the Witt algebra} or  {\it the classical Virasoro
algebra} (${\cal L}_{0} $).

If $\mathcal{G}$ is the algebra associated to the group $G,$
then:

\begin{eqnarray}
\cal{G} &=& \underbrace{{\cal L}_{0}} \oplus \underbrace{{\cal L}_{0}}.  \\
       & &  \mbox{{\small Witt}}    \hspace{0.4cm}  \mbox{{\small Witt}}
\nonumber
\end{eqnarray}

Infinitesimal transformations of the group $\Gamma$ are: 
$z\rightarrow z + \varepsilon(z),$
where $\varepsilon(z) $ is a infinitesimal analytical function. It
can be represented as an infinite Laurent series, viz.,

\begin{equation}
\varepsilon(z) =  \sum_n \varepsilon_{n} z^{n+1},  
\quad n\;\epsilon \; \mathbf{z}.
\end{equation}
Therefore, the Lie algebra ${\cal L}_{0}$ of the $\Gamma$
coincides with the algebra of differential operators defined in 
$\bf{C}- \{0 \}:$

\begin{equation}
\ell_{n}= z^{n+1} \frac{d}{dz},   \quad n=0, \pm 1, \pm 2, \cdots .
\end{equation}

The commutation relations have the following form:
\begin{equation}
 [ \ell_{n}, \ell_{m} ] = (n-m) \ell_{n+m}.
  \label{vir1}
\end{equation}
We shall denote the algebra (\ref{vir1}) as ${\cal L}_{0},$
which admits a unique 1-dimensional central extension:

\begin{equation}
{\cal L}_{\kappa} = {\cal L}_{0} \oplus \kappa \;\; 
\mbox{(The Virasoro algebra)},
\end{equation}
with the following commutation relations

\begin{eqnarray}
\relax [ \ell_{n}, \kappa ] & = & 0 \nonumber \\
\relax [ \ell_{n}, \ell_{m}  ] & = & (n- m) \ell_{n+m} +  \kappa
\frac{m^{3} - m}{12} {\delta}_{n+m , 0},
\end{eqnarray}
where the value of the  central charge $``\kappa''$ is the
parameter of the theory into the context of Quantum Field Theory.
The generators $ \ell_{-1}, \ell_{0}, \ell_{1} $ form the subalgebra $
sl(2,{\bf{R}}) \subset {\cal L}_{0}.$

In this subsection, we consider the oscillatory  representation
of the  elements for the  Virasoro algebra without modification.
Using the canonical commutation relation of the quantum mechanics
(\ref{RCC}), we obtain the following  commutador:

\begin{equation}
\label{cn}
[(a^-)^{n+1}, a^+]_- = (n+1)(a^-)^n, \quad n=0, 1, 2, \cdots .
\end{equation}
From this commutator, we see that the Virasoro operators
for the unidimensional harmonic oscillator can be defined by

\begin{equation}
\label{OV}
L_n=\left\{
\begin{array}{cc}
(a^-)^{n+1}a^+,\\  
a^-(a^+)^{n+1},
\end{array}
\right.
\end{equation}
which satisfy the following Virasoro algebra:

\begin{equation}
\label{AV}
[L_n, L_m]_- = (n-m)L_{n+m},
\end{equation}
where  $n\geq 0$ and  $m\geq 0.$
Next, we consider the oscillatory representation in terms of the 
Wigner oscillator.

\subsection{ Deformed Virasoro algebra}

Let us now consider the modified Virasoro algebra in terms of new 
ladder operators which satisfy the 
generalized commutation relation given by 
Eq. (\ref{CW}). Indeed, considering $L_n=(a^-)^{n+1}a^+$,
we see that the Eq. (\ref{cn}) becomes:

\begin{equation}
\label{cnWi}
[(a^-)^{n+1}, a^+]_- = \left\{
\begin{array}{cc}
(n+1)(a^-)^n+cP(a^-)^n, \quad n=2k  \\
 (n+1)(a^-)^n,\quad n=2k+1,
\end{array}\right.
\end{equation}
where $k=0, 1, 2, 3\cdots$.
Now we investigate the three possible cases for the Virasoro algebra.

{\noindent Caso (i): Two even indexes.}

In this case, the Virasoro algebra is not changed, i.e.

\begin{equation}
\label{AVpp}
[L_{2n}, L_{2m}]_- = 2(n-m)L_{2n+2m}.
\end{equation}

{\noindent Caso (ii): Two odd indexes.}

In this case, the Virasoro algebra is not also changed, i.e.

\begin{equation}
\label{AVii}
[L_{2n+1}, L_{2m+1}]_- = 2(n-m)L_{2n+2m+2}.
\end{equation}

Caso (iii): One even index and one odd index and vice-verse.

In this case, the Virasoro algebra is changed, i.e.,

\begin{equation}
\label{AVpi}
[L_{2n}, L_{2m+1}]_- = 2(n-m)L_{2n+2m+1}-(1-cP)L_{2n+2m+1}.
\end{equation}
Note that we get an anomalous term containng the parity operator $P$.
Besides, we can obtain the three possible cases
for the  Virasoro adjoint operators $L_n^{\dagger}=a^-(a^+)^{n+1}.$

The question we formulate now is the following: What is the  behaviour of the 
Virasoro operator on the autokets of the Wigner oscillator quantum states? 
To answer this question, one must first note that the Wigner 
oscillator ladder operators on autokets of these quantum states are given by

\begin{eqnarray}
\label{OW}
a^-_{\frac c2}|2m, \frac c2>&&=\sqrt{2m}|2m-1, \frac c2>\nonumber\\
a^-_{\frac c2}|2m+1, \frac c2>&&=\sqrt{2(m+E^{(0)})}|2m, \frac c2>\nonumber\\
a^+_{\frac c2}|2m, \frac c2>&&=\sqrt{2(m+E^{(0)})}|2m+1, \frac c2>\nonumber\\
a^+_{\frac c2}|2m+1, \frac c2>&&=\sqrt{2(m+1)}|2m+2, \frac c2>.
\end{eqnarray}
A detailed analysis on this question will appear in a forthcoming paper. 

\section{Conclusion}

In this work, we analyze  the Wigner-Heisenberg algebra to bosonic 
systems in connection with oscillators and, thus, 
we find a new representation for 
the Virasoro algebra. Acting 
 the annihilation operator(creation operator) in the
Fock basis $\mid 2m+1, \frac c2>(\mid 2m, \frac c2>)$
the eigenvalue of the ground state
of the Wigner oscillator appears only in the excited 
states associated with the even(odd) quanta given by Eq.(\ref{OW}).
We show that only in the case associated with one even index
and one odd index in the operator $L_n$
the Virasoro algebra is changed. 

The super-realization of the  Wigner-Heisenberg algebra proposed
by Jayaraman and  Rodrigues \cite{JR90}, and independently by
Plyushchay \cite{Mi00}, has been changed to 
investigate a potential model that describes a hydrogen 
atom with para-statistics \cite{Bec93}.

\vspace{1cm}


\centerline{\bf ACKNOWLEDGMENTS}

The authors are grateful to the J. A.
Helayel Neto for the kind hospitality at CBPF-MCT and 
for fruitful discussions. 
RLR and HLC are grateful to the CNPq - Brazil for their post-doctoral 
and doctoral Graduate fellowships,
respectively. The authors would also like to thank Plyushchay
and Zachos
for the kind interest in pointing
out relevant references on the subject of this paper.
This work was presented in the XXII Brazilian National
Meeting on Particles and Fields (October/2001).


\begin{thebibliography}{99}


\bibitem{Wigner50} Wigner E P 1950 {\it Phys. Rev.} {\bf 77} 711

\bibitem{Yang} Yang L M 1951
{\it Phys. Rev.} {\bf 84} 788

\bibitem{Ohnuki} Ohnuki Y and Kamefuchi S 1978 {\it J. Math. Phys.}
{\bf 19} 67; Ohnuki Y and  Watanabe S
1992 {\it J. Math. Phys.} {\bf 33} 3653

\bibitem{Sharma} Sharma J K,  Mehta C L, Mukunda N and Sudarshan E C G
1981 {\it J. Math. Phys.} {\bf 22} 78; Mukunda N, Sudarshan E C G,
Sharma J K and Mehta C L 1980 {\it J. Math. Phys.} {\bf 21} 2386;
Sharma J K, Mehta C L and
Sudarshan E C G 1978 {\it J. Math. Phys.} {\bf 19} 2089

\bibitem{Green} Green H S 1953 {\it Phys. Rev.} {\bf 90} 270;
 Macfarlane A J 1994 {\it J. Math. Phys.} {\bf 35} 1054

\bibitem{Ryan} O'Raifeartaigh L and Ryan C 1963
{\it Proc. R. Irish Acad.} {\bf A62} 93

\bibitem{Deser} Boulware D G and Deser S 1963 {\it Il Nuovo Cimento}
{\bf 230XXX} 230

\bibitem{JR90} Jayaraman J and Rodrigues R de L
1990 {\it J. Phys. A: Math. Gen. }
{\bf 23} 3123

\bibitem{Mi97} Plyushchay M S 1997 {\it Nucl. Phys.}
{\bf B491} 619 

\bibitem{Mi00} Plyushchay M S 2000 {\it Int. J. of Mod. Phys.}
{\bf A15} 3679 

\bibitem{Top} Toppan F, ``Conformal Field Theories,''
preprint UCP-HEP-90/7; Belavin A A, Polyakov A M, Zamoloochikov A B,
1984 {\it Nucl. Phys.} {\bf B241} 333

\bibitem{zac87}
Fairlie D B and Zachos C K,
1988 {\it Phys. Lett.} {\bf B202} 320

\bibitem{zac88} Fairlie D B, Nuyts J and Zachos C K,
1988, {\it Commun. Math.Phys.} {\bf 117} 595 

\bibitem{zac90} Curtright T L, 
Cosmas K. Zachos C K, 1990 {\it Phys. Lett.} {\bf B243} 237


\bibitem{Bec93} Beckers J and Debergh N 1993 {\it Phys. Lett.}
{\bf A178} 43

\end{thebibliography}
\end{document}